# Paper-Patent Citation Linkages as Early Signs for Predicting Delayed Recognized Knowledge: Macro and Micro Evidence

Jian Du[1,2], Peixin Li[2], Robin Haunschild[3], Yinan Sun[2], Xiaoli Tang[2]*

1 National Institute of Health Data Science, Peking University, Beijing, 100071 (China)
2 Institute of Medical Information & Library, Chinese Academy of Medical Sciences, Beijing, 100005 (China)
3 Max Planck Institute for Solid State Research, Stuttgart, 70569 (Gemany)

**Abstract**
In this study, we investigate the extent to which patent citations to papers can serve as early signs for predicting delayed recognized knowledge in science using a comparative study with a control group, i.e., instant recognition papers. We identify the two opposite groups of papers by the *Bcp* measure, a parameter-free index for identifying papers which were recognized with delay. We provide a macro (Science/Nature papers dataset) and micro (a case chosen from the dataset) evidence on paper-patent citation linkages as early signs for predicting delayed recognized knowledge in science. It appears that papers with delayed recognition show a stronger and longer technical impact than instant recognition papers. We provide indication that in the more recent years papers with delayed recognition are awakened more often and earlier by a patent rather than by a scientific paper (also called "prince"). We also found that patent citations seem to play an important role to avoid instant recognition papers to level off or to become a so called "flash in the pan", i.e., instant recognition. It also appears that the sleeping beauties may firstly encounter negative citations and then patent citations and finally get widely recognized. In contrast to the two focused fields (biology and chemistry) for instant recognition papers, delayed recognition papers are rather evenly distributed in biology, chemistry, psychology, geology, materials science, and physics. We discovered several pairs of "science sleeping"-"technology inducing", such as "biology-biotechnology/pharmaceuticals", "chemistry-chemical engineering", as well as some trans-fields science-technology interactions, such as "psychology - computer technology/control technology/audio-visual technology", "physics - computer technology", and "mathematics-computer technology". We propose in further research to discover the potential ahead of time and transformative research by using citation delay analysis, patent & NPL analysis, and citation context analysis.

**Keywords:**
Delayed recognition papers; Citation delay analysis; Patent & NPL analysis; Scientific knowledge; Early signs

## 1 Introduction

According to our understanding of Kuhn's paradigm on the *Structure of Scientific Revolutions*, scientific knowledge often develops incrementally (incremental research), occasionally disrupted by paradigm-shifting discoveries (transformative research) (Kuhn & Hawkins, 1963). In contrast to incremental research, which moves forward through the continuous, incremental accumulation of knowledge, transformative research drives science forward by radically changing our understanding of a concept, causing a paradigm shift, or opening new frontiers (Trevors, Pollack, Saier, & Masson, 2012). Prioritization of transformative research has become pervasive among funding agencies (Sen, 2017). Such research brings great rewards, but also carries great risks for funding agencies because transformative research projects are very hard to identify in their early stages. An on-going challenge lies in identifying transformative research projects at the time they are proposed. Although it is almost impossible to predict the transformative potential and effect of research during the proposal stage, yet it is more predictable during the research process or even for a long time after the discovery. Further, transformative research should not be understood as just the opposite of incremental research. Actually, most transformative research began with incremental goals, and the transformative effect was recognized later (Gravem et al., 2017).

Ahead-of-time knowledge and transformative research play key roles for the development of science, but they are often at the very beginning ignored or resisted by the scientific community and thus are subjected to delayed recognition (Figure 1). In a report by the National Academies of Sciences, Engineering, and Medicine in 2016, the authors reviewed five transformative research areas of geographical science that have taken shape over the past 65 years to explore how transformative research has emerged. They found that transformative innovations can arise from older and long-ignored ideas (National Academies of Sciences & Medicine, 2016). Such ideas are often called "Sleeping Beauties" (SB) in science, one type of publications that rarely cited (or "sleeps") for a long time and then, almost suddenly, attracts a lot of citations (or "is awakened by a Prince") (Van Raan, 2004). This concept, in terms of citation curve of a given paper, actually quantifies the phenomenon of "delayed recognition of scientific achievements" (Hook, 2002). To the best of our knowledge sociologist Stephen Cole was the first to propose one could use citations to measure delayed recognition in science (Cole, 1970).

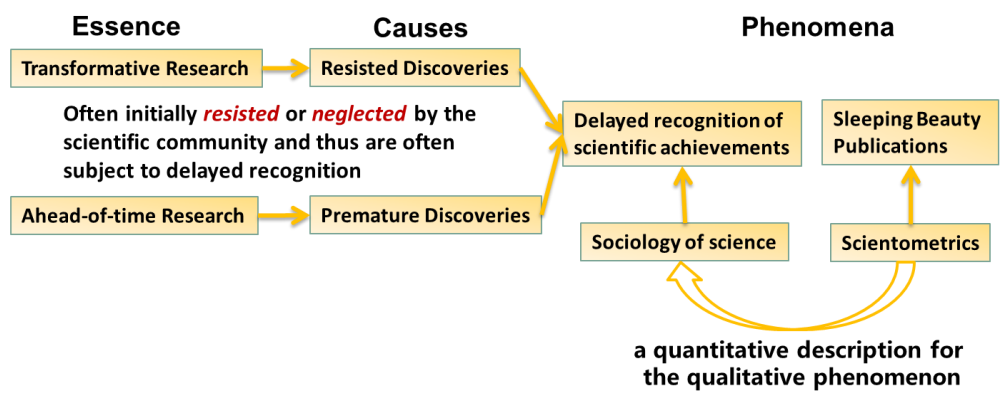

**Figure 1. A schematic model of sleeping beauty and delayed recognition in science**

Among the many papers written by colleagues on delayed recognition or SBs which have been discussed in our foregoing papers (Du & Wu, 2016, 2018), we concentrate on a few important recent developments and focus on the early identification of such type of publications and/or transformative research. Can we know in early stages if a research project looks promising or might lead to transformative research? The studies mentioned below may provide some insight into early signs of the awakening of SB publications or the recognition of premature discoveries and transformative research. It has been shown by Marx (2014) using the example of the paper by Shockley and Queisser (1961) that delayed recognition papers often start getting cited when a highly cited paper or a well-known author has paid attention to them. But it is difficult to define such highly-cited paper or prominent scholar just at that time.

One type of citation-based early signs focused on trans-disciplinary citing pattern. Dey, Roy, Chakraborty, and Ghosh (2017) analysed the features of a given paper which may become an SB, and were the first to investigate the early identification of SBs in computer science. They developed a methodology to early predict whether a paper is likely to become an SB. They observed that the potential of a given paper to receive citations from multiple fields is the most important feature. This observation is corroborated by Ke, Ferrara, Radicchi, and Flammini (2015). They found that in many cases, the awakening of SBs occurs when an application in a field outside of the SB's field is found, such as statistical methods that became useful in biology, chemistry, or physics.

Another type of citation-based early signs focused on patent citations to papers, i.e., the correlation between the SBs and the scientific non-patent references (SNPR) in patents. An established belief is that SBs relate to more basic and theoretical concept rather than more applied and experimental work. But the fact is that half of the SBs in physics, chemistry, engineering, and computer science are applied research or published in application-oriented journals (Van Raan, 2015) and significantly more cited by patents than other common scientific papers (van Raan, 2017). The time lag between the publication year of the SBs and their first citation in a patent becomes shorter in recent years (van Raan, 2017). van Raan & Winnink (2018) investigated this further by using younger SBs cited in patents. Their observations suggest that, on average SBs are awakened increasingly earlier by a patent ("technological prince") rather than by a scholarly paper ("scientific prince") in the more recent years. Very recently, the scientific and technological impact of sleeping beauties in medical research fields was analysed by van Raan and Winnink (2019). Du and Wu (2018) also found that 60% of the extreme SBs published in Science/Nature have technological impact (cited by patents), and their first citation in a patent is usually earlier than the awakening year.

But findings by van Raan (2017) and van Raan and Winnink (2018) on a more density of patent citations to sleeping beauty publications were drawn by comparing those with all Web of Science publications, rather than with the just the opposite group, for example the instant recognition papers used in our study. In their foregoing papers, they analyzed a set of SBs with such thresholds as: (1) the average number of annual citations is at most one during 10 years after publication and, (2) the average annual citations is at least five during the next 10 years after 10 years of sleep. So, we can expect that the SBs have been cited at least 50 times. They identified 389 SBs for physics, 265 SBs for chemistry, and 367 SBs for engineering and computer science and found that 62 (16%), 92 (35%), and 108 (29%) of those SBs are also cited by patents. The possibility of SB papers being cited by patents is obviously higher than the proportion of all Web of Science (4.4%) or MEDLINE covered publications cited by patents (4%) according to Ahmadpoor and Jones (2017) and (Ke, 2018). The percentage 4% is calculated based on all papers indexed by Web of Science/MEDLINE, including the papers which have never been cited. According to an investigation on the uncited papers in Web of Science (Van Noorden, 2017), basically about one in fifth (21%) publications across all disciplines during 1900-2015 haven't been cited. Setting papers published in 2006 as the sample, the proportion of uncited papers within ten years after publication varies across different disciplines, for example 4% for biomedical science, 8% for chemistry, 11% for physics, and 24% for engineering and technology. Since these papers have no scientific impact, they are likely to have no technical impact and thus will probably not be cited by patents.

In this study, we will answer two questions based on van Raan's work: (1) to what extent patent citations to papers can serve as early signs of delayed recognition using a comparative study with a control group, i.e., instant recognition papers? Delayed recognition in science is a phenomenon where papers went unnoticed until they are re-discovered some years after publication. By contrast, instant recognition (also called "flashes in the pan") in science is a phenomenon where papers received a lot of citations shortly after publication, but were ignored very quickly (Ye & Bornmann, 2018). (2) What is the pattern of science-technology interaction between the sleeping science and the technology inducing its recognition? Based on our previous investigations on systematic identification of SBs and on their awaking mechanisms (Du & Wu, 2016, 2018), this contribution will further validate and detect early signs of the awakening of SBs.

## 2 Data and Methods

To characterize delayed recognition papers, it is necessary to compare them with instant recognition papers. We will turn an apparent yes/no question into a continuous phenomenon.

*2.1. A parameter-free index for measuring the extent of delayed recognition*

In our foregoing paper, we introduced a parameter-free index to measure the extent of citation delay based on yearly cumulative percentage of citations for any given paper Du and Wu (2018). The index is denoted as Beauty coefficient based on cumulative percentage of citations, *Bcp* for short. *Bcp* depends on the shape of the cumulative citation curve, especially when there is a cumulative citation burst in the whole citation life cycle for a given paper. *Bcp* allows comparing the extent of delayed citation impact for publications in different fields with different amount of citations since it considers yearly cumulative percentage of citations.

The size of the analysis window (the number of years during which citations are observed) does have an influence on the value of *Bcp*. Basically, the larger the analysis window for a given paper, the lager value of *Bcp* is to be observed. It is similar to the case that older papers are potentially more cited than new papers.

In general, it is a sign of delayed recognition if the cumulative citation curve for a given paper is concave (*Bcp*>0). Instant recognition is indicated by a convex cumulative citation curve (*Bcp*<0). The larger the *Bcp* value, the more delayed recognition of a paper in terms of the citation curve. The smaller the *Bcp* value (negative value), the more instant recognition of a given paper. Just like the "top 1%" is usually used to select highly cited papers, we will also use "top 1%" versus "bottom 1%" for grouping delayed recognition and instant recognition papers.

The awakening year means the time when the abrupt change of the cumulative citations, not the annual citations from one given year to the next year occurs. As is shown in Figure 2, $t$ is the time interval after publication for a given paper and $C_t$ the cumulative percentage of citations within the observing window, $C_t \in [0,1]$. We define $t_m$ the last year of the citation window, namely the year that the percentage of citations accumulated to 100%. The reference line $y_t$ connects the point (0, $C_0$) and ($t_m$, 1). We define the abrupt change as the time $t$ at which the distance $d(t)$ between the point (t, $C_t$) and the reference line $y_t$ reaches its maximum. So, generally when the cumulative citation curve for a given paper is concave (*Bcp* > 0), the time $t_a$ can be called "awakening time" and the citations started to show an abrupt increase. In contrast, when a given paper's cumulative citation curve is convex (*Bcp* < 0), the time $t_f$ can be called "falling time" and the citations started to show an abrupt decrease. So, both the awakening year and the falling year are calculated from the whole life cycle of citations for a given paper. Just like that the awakening year is not always the year when the number of yearly citations is at the minimum, the falling year is also not always the year when the yearly citations is at the maximum. The definition of the falling year is just the same with the awakening year in a unified framework.

Figure 2 illustrated the definition of the awakening year and falling year in a unified *Bcp* framework. The awakening for the delayed recognized Nature article (*Bcp* 15.024) is 2004,

Whereas the falling year of the instantly recognized Science article (*Bcp* -18.609) is 1977. The awakening year and falling year are in accordance with the annual citation profile shown in Figure 3 A and B.

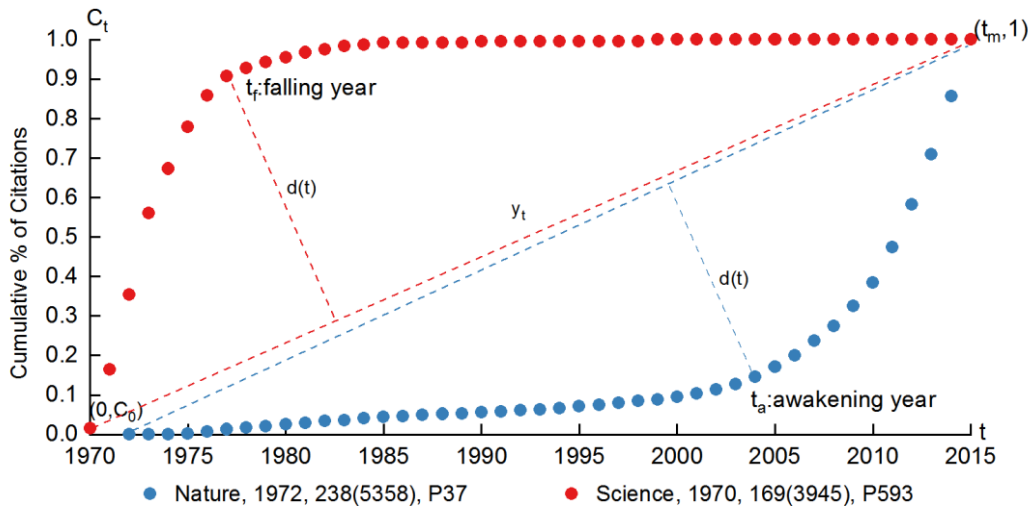

**Figure 2. Definition of the awakening year and falling year in a unified *Bcp* framework.**

*2.2. Delayed recognition (top 1%) versus instant recognition (bottom 1%) papers*

Both concepts of "delayed recognition" (DR) and "instant recognition" (IR) contain the meaning of "recognition". How to represent "recognition"? We use the number of citations to describe the extent to which a given paper attracts attention and/or gets recognition from the scientific community at the observing year. So, we need to identify DR and IR papers from those received a certain amount of citations. But it is very difficult to define such a lower threshold of citations.

In addition to a description for "recognition", it is necessary to consider "delay" when collecting the dataset. In order to make sure each paper in our data set has at least 10 years to accumulate citations, articles published in Science and Nature during 1970-2005 are included, and the citation data for these articles are collected till the end of 2015. The citation window is 10 years longer than the publication window. Thus, 2005 is the most recent publication year included in our study. Set the year 2015 as the observed ending point, there are 78,403 articles published during 1975-2005, of which those with at least 200 citations, in total 20,000 publications are ready for the following analysis. We selected the top 1% (N=200) as delayed recognition papers and the bottom 1% (N=200) as instant recognition papers.

The important reason for selecting publications in Science & Nature is the multidisciplinary nature of such journals, where more potential delayed recognition papers exist Ke et al. (2015).. Science and Nature are the most well-known multi-disciplinary journals, which always attract the attention from all academic communities instead of specific fields. Such venues not only publish trending topics-which may obsolete soon and become flashes in the pan-but also publish high quality contributions which are too premature to be understood, futuristic predictions, silent for a long time, and will not revive until they are verified by the academic community. Thus, we believe that the Science/Nature papers dataset is a good corpus to mine the knowledge of delayed recognition and instant recognition.

*2.3. Patent-citing related indicators*

Patent documents provide backward citations to earlier patents and non-patent literature (NPL), which includes peer-reviewed journal articles and other documents. Using the patent backward citations to NPL, one can measure the technological impact of the scientific knowledge (Roach & Cohen, 2013). We compare the extent to which the delayed recognition papers and the instant recognition papers show up as NPL in patents. In this study, the data of NPLs cited by patents was gathered by searching Lens.org, a digital platform linking the patent documents to the scholarly literature across the world through collaborations with CrossRef and National Library of Medicine (Jefferson et al., 2018). Based on the lens score, i.e., the number of publications cited by patents, the Nature Index published the top 200 innovative institutions in 2017. The innovation ranking opens a new window to observe the impact of academic research on technical innovation by examining how scientific articles are cited in third party patents. In this study, we mainly focused on the following indicators.

1) *Number of citing patent families*: to prevent double counting when quantifying the number of patent citations to a given paper. A patent family is a set of patents taken in various countries to protect a single invention (when a first application in a country – the priority – is then extended to other offices). In other words, a patent family is the same invention disclosed by a common inventor(s) and patented in more than one country.
2) *Forward Citations of the Earliest Priority Patent*: whether they are firstly cited by an important patent.
3) *Durability of Patent Citing*: the interval of the filing years between the latest and the earliest priority patent. By this measure, we can figure out the durability of patent citing to a given paper.

*2.4. Fields of study*

In order to compare the field of technology of papers with fields of study, we use the hierarchical fields of study from Microsoft Academic which are provided by a semantic algorithm on the paper basis. Microsoft Academic is an unconventional bibliographic database in the sense that it does not receive the publication's meta-data from the publishers directly. Microsoft Academic's content is mainly determined from what the search engine Bing can find and identify as scientific publication on the internet. This introduces a bias towards scientific literature which is available on the internet. However, we can expect that the studied publication set (*Science* and *Nature* publications) is available on the internet. Traditional multi-disciplinary databases use journal sets for field classification. Such a classification scheme is not helpful for our current study. Thus, Microsoft Academic provides very important information for our study with the hierarchical fields of study based on a semantic algorithm on the paper basis. Furthermore, Microsoft Academic is one of the main data sources of Lens.org. Therefore, it is a natural choice to use both databases in this study. We appended the field of study from a local in-house database of Microsoft Academic (maintained at the Max Planck Institute for Solid State Research) to the top 1% and bottom 1% papers via the DOI and from Lens.org via PMID. Starting in August 2018, all scholarly papers cited by patents will have the information of field of study thanks to a partnership with Microsoft Academic. Not all papers in Microsoft Academic database have a field of study attached to them but some papers have multiple fields of study at different levels. We found at least one field of study for 198 top 1% papers and for 196 bottom 1% papers with DOIs and/or PMIDs. For the rest of papers, we give the top-level field of study based on their research areas reflected by title and/or abstract.

## 3 Results from a comparative study between delayed and instant recognized papers

*3.1 Identifying the two opposite groups of papers by Bcp measure*

Figure 3 shows citation curves of the first and the last paper ranked by *Bcp* and the distribution of citation percentiles for the two groups of papers. The awakening year for the most one delayed recognition paper is 2004 (until the 33$^{rd}$ year after publication) and the falling year for the most one instant recognition paper is 1977 (just in the 7$^{th}$ year after publication). In Figure 3C and Figure 3D, the percentile is based on all 20000 papers in Nature and Science included in our analysis. We ranked all the included 20,000 articles by *Bcp* measure and the number of citations, respectively. We can see many of the most delayed recognition papers (top 1% by *Bcp* measure) are lowly cited, whereas many of the most instant recognition papers (bottom 1% by *Bcp* measure) are highly cited. We can conclude that *Bcp* is not very dependent on the total number of citations of a given paper.

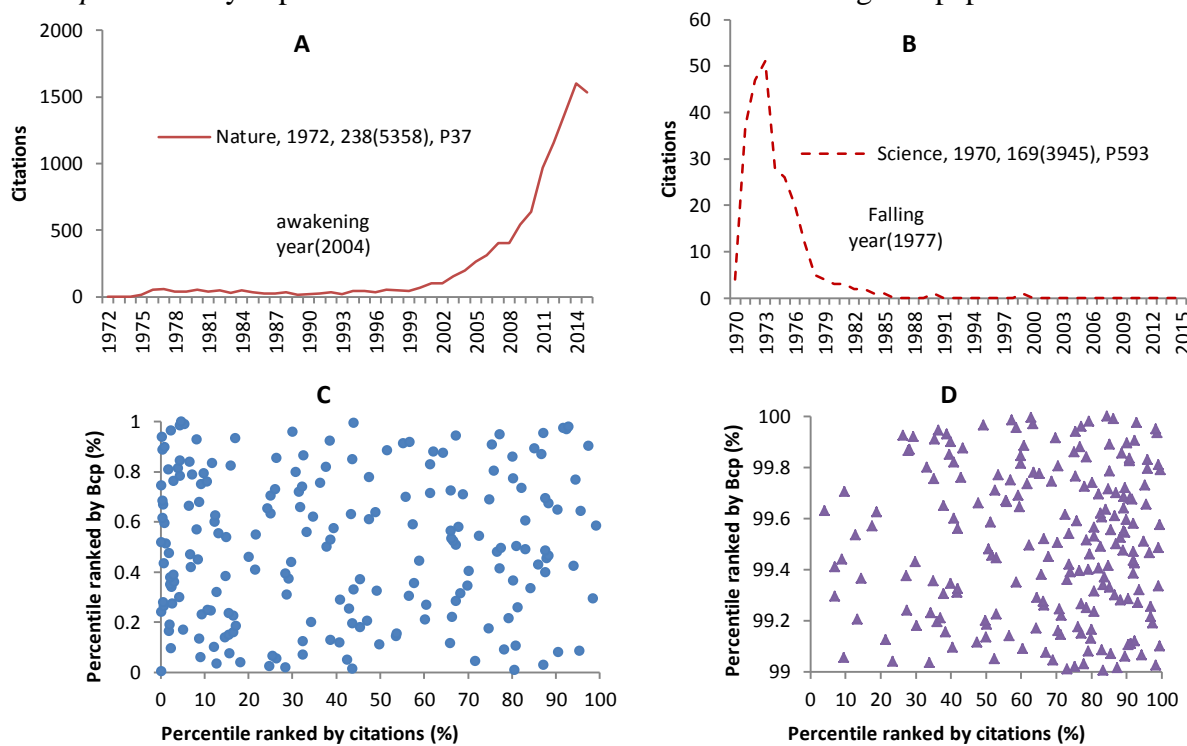

**Figure 3. Annual citations of the first (A) and the last (B) paper ranked by *Bcp* and distribution of citation percentiles for top 1% (C) and bottom 1% (D) papers**

*3.2 Delayed recognition papers showing a stronger and longer technical impact than instant recognition papers*

We find that about half of (99/200) the delayed recognition papers are cited by patents (DR-NPLs) and about one-third of (70/200) the instant recognition papers are cited by patents (IR-NPLs). Similar to citations given by scientific publications, the number of citations by patents is characterized by a skewed distribution. For example, about one third of the DR-NPLs are cited by only 1 or 2 patents, and six DR-NPLs are cited by more than 300 patents. In addition, about half of the IR-NPLs are cited by only 1 or 2 patents. One IR-NPLs is cited by 120 patents, but the rest is cited by no more than 40 patents. In total, the 99 DR-NPLs are cited by 3988 patents, and the 70 IR-NPLs are cited by 543 patents.

The citations and patent citations for the two groups of papers do not follow a normal distribution, skewed distributions should not be studied in terms of central tendency statistics such as the average number of citations, but rather using nonparametric statistics, such as rate ratio. Since both the two group have the sample paper which have not been cited by patents, the rate ratio was used to compare the patent citing profile between the two groups. The rate ratio tells us how more common a particular event, i.e., cited by patents, happened in the observed group of DR papers than in the control group of IR papers.

Table 1. DR versus IR papers: showing up as NPLs in patents

|  | Group | Yes | No | Rate | 95% Confidence Interval | $P$ |
|---|---|---|---|---|---|---|
| Citing Patent Families | DR | 99 | 101 | 0.495 | (0.426,0.564) | 0.003 |
|  | IR | 70 | 130 | 0.350 | (0.284,0.416) |  |
| Forward Citations of the Earliest Priority Patent | DR | 82 | 118 | 0.41 | (0.342,0.478) | 0.009 |
|  | IR | 57 | 143 | 0.285 | (0.222,0.348) |  |
| Durability of Patent Citing | DR | 75 | 125 | 0.375 | (0.308,0.442) | <0.001 |
|  | IR | 41 | 159 | 0.205 | (0.149,0.261) |  |

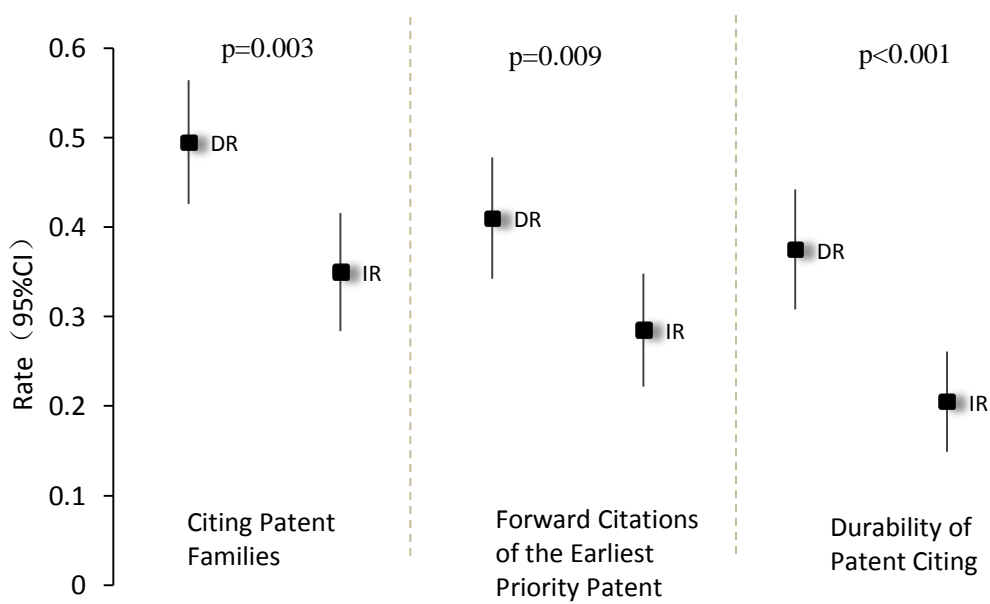

**Figure 4. DR versus IR papers: showing up as NPLs in patents**

Table 1 and Figure 4 show DR are more often cited by patents than IR, and DR are cited by more important patents than IR papers, showing a stronger technical impact. Next, we determined for the DR-NPLs and IR-NPLs the filing years of the earliest and of the latest citing patent until Dec 23, 2018. The difference between the filing years of the latest and the earliest patent indicates the durability of patent citations. DR papers have a longer technical impact than IR.

*3.3. Patent citing is earlier than awakening in the more recent years*

First, we compare the earliest patent citing year and the awakening year for the 99 delayed recognition NPLs, and find that for 70% (n=69) of the papers the first patent citing year is earlier than the awakening year; for 5% (n=5) of the papers the first patent citing year is the same as the awakening year; only 25% (n=25) of the papers are cited by patents after awakening (Figure 5A).

The difference between the publication year and the filing year of the earliest citing patent defines how soon a given paper subsequently be cited by a patent. We firstly calculated the time lag between publication year and the first patent citing year for each of the 99 DR-NPLs. This time lag ranges from 0 to 41 years (median 14, SD=10.9). Time lag with zero means the article is cited by a patent just at the publication year. In order to figure out if there is a trend over time, we calculate the averages for successive, partly overlapping 5-year periods (Figure 5B). In the case of the 99 DR-NPLs, these periods are 1970-1974, 1971-1975, …, and 1990-1994. Remarkably, the time lag becomes rapidly shorter in the more recent years within the observed period 1970-1994. In other words, for the more recent published DR-NPL, it is more quickly cited by a patent.

Then, we counted the average time lag between the first patent citing year and the falling year for instant recognition papers. The difference between the filing year of the earliest citing patent and the falling year define the time lag between the first patent citation and its "falling". The "falling" means the citations of the IR-NPL begin to abruptly decrease. This time lag ranges from -28 to 16 years (median 4.5, SD=10.8). For example, for a Science article published in 1975 (with DOI 10.1126/science.1094538), the falling year is 1982, and the year of the first patent citation is 1980. So, time lag between the year of the first patent citation and awakening year is 2. In order to investigate if there is a trend over time, we also calculate the averages for successive, partly overlapping 5-year periods. Obviously, the time lag becomes rapidly longer in the measured period 1970-1983 (Figure 5C). In the more recent years, even the instant recognition papers will be more likely to exhibit a longer durability of citations once they are cited by a patent. Both observations suggest that, on average, in the more recent years, the delayed recognition papers are awakened increasingly earlier by a patent ("technological prince") rather than by a scholarly paper ("scientific prince"). Patent citations seem to play an important role to avoid instant recognition papers to level off or become "flashes in the pan".

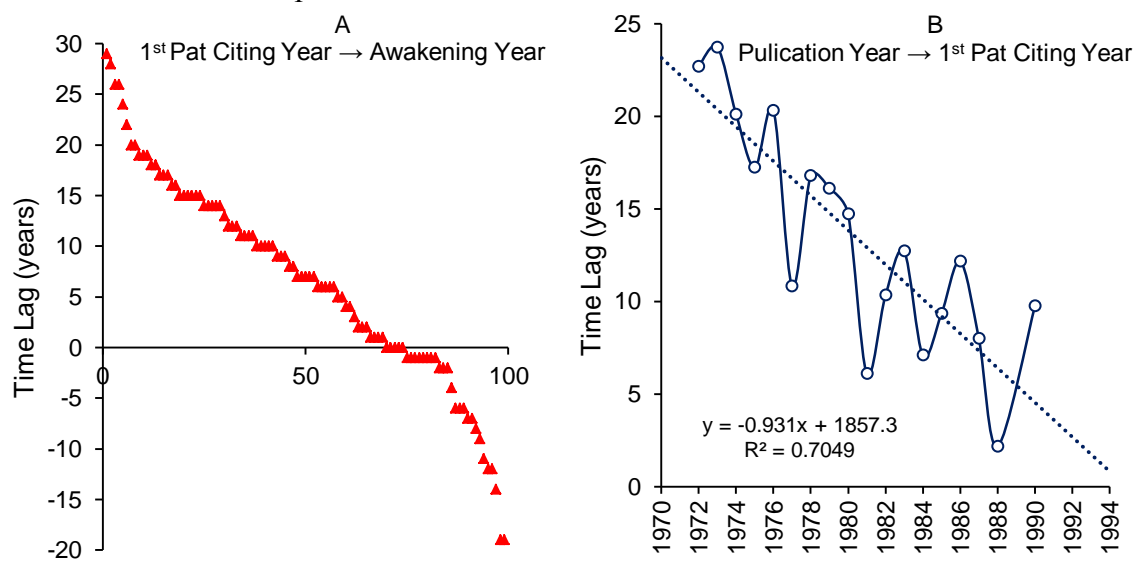

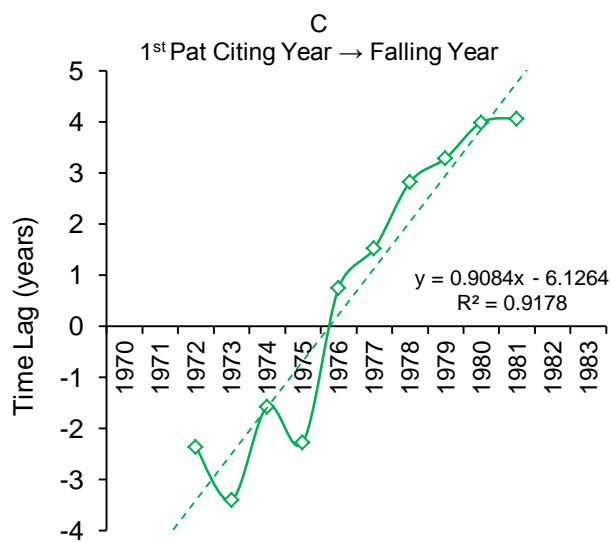

**Figure 5.** Time lag between the earliest patent citing year and the awakening year for 99 DR-NPLs (A), and the trend of time lag between publication year and 1st patent citing year for 99 DR-NPLs (B) and the trend of time lag between 1st patent citing year and the falling year (when the citations of the IR-NPL begin to abruptly decrease measured by our *Bcp* framework) for 70 IR-NPLs (C).

*3.4 Comparing the difference of science-technology interactions between DR and IR papers*

It has been argued that technology-driven, or more specific, patent citations to papers, might be one of the awakening mechanisms for delayed recognition papers (Du, Sun, Zhang, & Tang, 2019). To reveal the structure of research fields for the scientific papers and the interactions with the technical focus of the citing patent families, we firstly match the field of study for each of the 200 delayed recognition and the other 200 instant recognition papers from Microsoft Academic, which determines the field of study based on machine learning parsing of all accessible text in a record. Microsoft Academic increases the power of semantic search by adding more fields of study (from February 15, 2018). There are now 19 top-level fields of study, including biology, medicine, geology, chemistry, psychology, philosophy, sociology, engineering, economics, computer science, art, physics, history, political science, materials science, mathematics, geography, business, and environmental science. The Microsoft Academic data contain fields of study with a six-level hierarchy. Using the technology classification groups of WIPO concordance table[1], which links IPC symbols with 35 fields of technology we identified the fields of technology for each of the earliest citing patents in our two datasets. Afterwards, we map the interactions (by means of direct citations) between fields of study and fields of technology to figure out the different patterns for the two groups of papers.

There are 952 fields of study for the 200 delayed recognition papers, of which 55 (27.5%) are biology, followed by chemistry, psychology (n=25, 12.5%), geology (n=17, 8.5%), materials science (n=17, 8.5%), physics (n=16, 8%), and so on. However, there are 618 fields of study for the 200 instant recognition papers, of which almost 90% (n=180) are biology and nearly 10% (n=19) are chemistry. Figure 6 shows that delayed recognition papers in biology are mainly cited by patents in biotechnology and pharmaceuticals. Delayed recognition papers in chemistry are often cited by chemical engineering technology, biotechnology, and pharmaceuticals. Delayed recognition papers in materials science are mainly cited by patents in biotechnology and metallurgy materials. Delayed recognition papers in psychology are mainly cited by computer technology and control technology. Delayed recognition papers in physics are mainly cited by computer technology.

---

[1] https://www.wipo.int/export/sites/www/ipstats/en/statistics/patents/pdf/wipo_ipc_technology.pdf

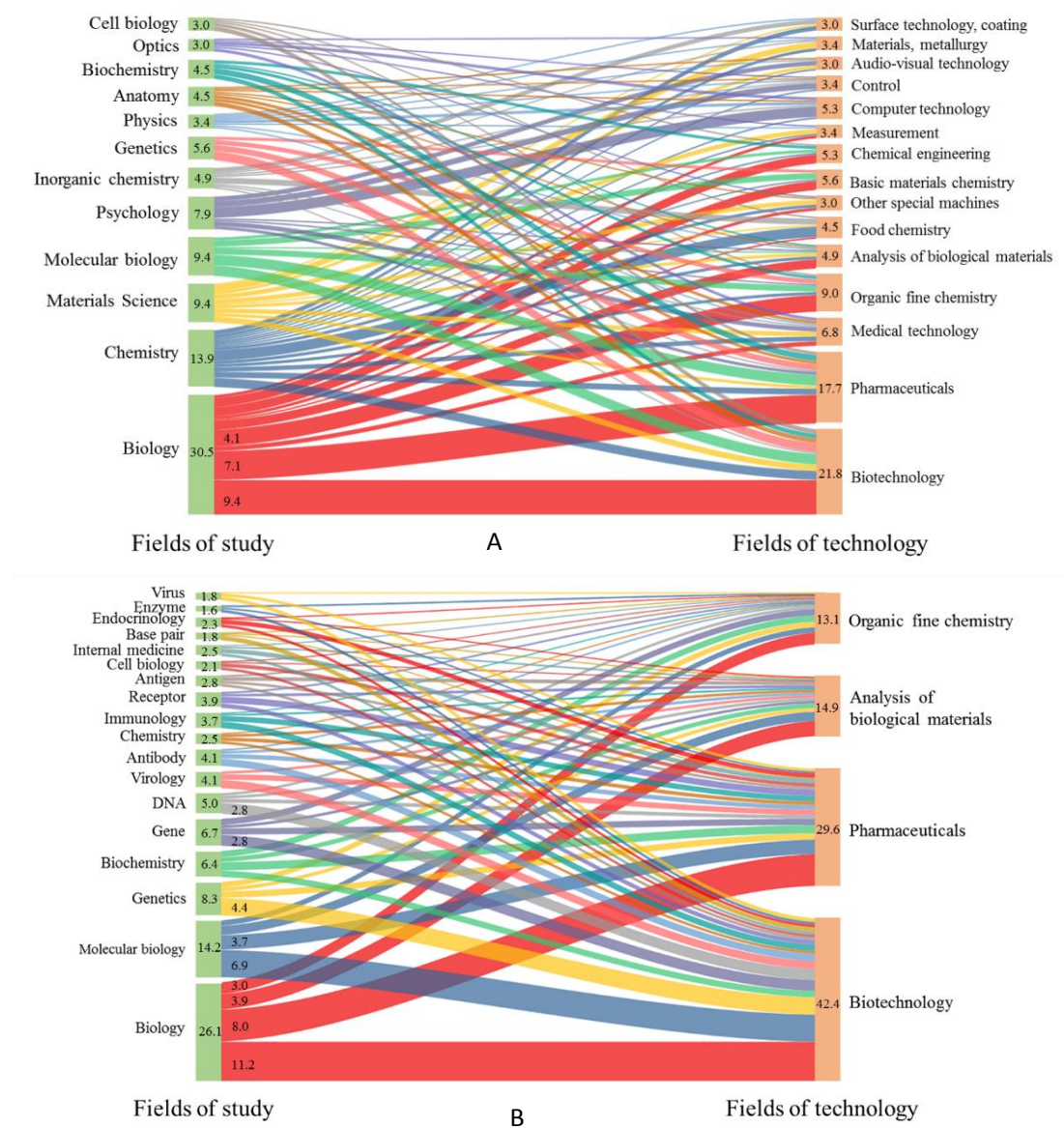

**Figure 6. Interactions between fields of study of papers and their earliest citing patent families' fields of technology for delayed recognition papers (A) and Instant recognition papers (B).**

**4 A case study on John Maynard Smith's paper about the concept of protein space**

*4.1 Comparison of increase of citations with growth of the research field*

From the top 200 delayed recognition papers, we find a case on John Maynard Smith's ahead of time concept of protein space, which was commented by *Nature* that "1970 Foreshadows concepts now widely applied in studies of molecular evolution, such as genotype-phenotype mapping"[2]. Evidently, it is a typical case of ahead of time research.

---
[2] http://www.nature.com/nature/focus/maynardsmith/

We collected the citation data of John Maynard Smith's Nature article from the MA database. This article received 494 citations between 1970 and 2017. This results in an annual average growth rate (AAGR) relative to the awakening year (1987) of 44.68%. John Maynard Smith's paper is assigned to eight different fields of study in MA. The AAGRs of these eight fields of study are compared to the AAGR of Smith (1970) in Table 2. The AAGRs of the fields of study are all below 10% relative to 1987. This indicates that the increase in citations of the paper in our case study is only to a minor extent due to the overall growth of its research field.

**Table 2. Annual average growth rate relative to 1987 (the awakening year) for citations to Smith (1970) and publications assigned to the fields of study which Smith (1970) has been assigned to**

| Data set | AAGR [%] |
| --- | --- |
| Citations to Smith (1970) | 44.68 |
| Biology papers | 5.46 |
| Peptide sequence papers | 0.66 |
| Amino acid papers | 1.99 |
| Natural selection papers | 7.34 |
| Nucleotide papers | 0.66 |
| Abiogenesis papers | 8.10 |
| Genetics papers | 7.20 |
| Biochemistry papers | 2.80 |

*4.2. The earliest citations were notable negative and then obvious positive after ten years*

This paper was cited by three papers immediately after publication, and received only 10 citations within 15 years. In terms of citation context analysis, we found that the most two earliest citing papers evidently resisted or refuted Smith's concept, using such negative comment as the following sentences (Table 3).

| |
| --- |
| SMITH has compared protein evolution with a popular word game…I would point out, ***however***, that it is ***not sufficient*** to…Smith gives the impression that he is ***disagreeing with*** Salisbury…In conclusion, after ***examining*** Smith's argument, Salisbury's contention still seems to stand… |
| My particular doubt has been published (Salisbury, 1969), scientists have taken their ***shots at*** it (Smith, 1970), and it has been ***defended*** (Spetner, 1970). |

In fact, in his own paper, Smith described that "*Salisbury has argued that there is an apparent contradiction between two fundamental concepts of biology…Natural selection is therefore ineffective because it lacks the essential raw material —favorable mutations…I should like to look at the problem from a different point of view*" (Smith, 1970, p. 564). So, Smith's concept also has the meaning of transformative research which challenges established ideas and thus is subject to resistance or controversy shortly after publication.

But, in two citing articles published in 1982, the citing author directly commented that "the protein space ***concept first introduced by*** Maynard Smith (1970)", and "Maynard Smith ***proposed an interesting idea of a protein space***". They are evidently positive comments when citing the focal paper. For more information please refer to Appendix Table 1.

But, shortly after 1982 when obviously positive comments occurred, Smith (1970) did not attract a lot of citations suddenly due to the so fundamental and basic science concept. We then investigate is there any other triggering mechanism behind the subsequent increase of citations.

*4.3. The sleeping beauty was cited by patents*

We checked all the citing patents to Smith's SB paper via lens.org, and found that it was cited by two patent families, US5824469 (Priority date Jul 17, 1986) and WO2003075129 (Priority date Mar 1, 2002). The former patent has been filed by a university, and the latter one has been filed by a company (Table 3). The priority dates of the two citing patents were consistent with the citations increasing year according to the citation curve of Smith (1970). Looking at the citation curve of John Maynard Smith's paper, there is no strong increase in the citation profile. Instead, it has multiple peaks. At around 1986, the citations begin to increase and after the year 2002, the citations show the second increase. The concept of protein space became the basis of some modern biological technologies. For example, while the inventors of the first patent did not cite Smith (1970) in the patent full-text, the examiner included Smith (1970) in the list of cited references of the patent when assessing the novelty of technology since they are highly content related. Since it is not the inventor citing but the reviewer citing, there is no citing statement in the main text of the patent (Table 3). Based on the concept of protein space, the first patent developed a new method to screen certain cells exhibiting the predetermined biological function by inserting random nucleotide sequences. The inventors of the second patent put an important emphasis on Smith (1970). From the citation context (Table 3), we can see the second patent invented new ways to efficiently search sequence space to identify functional proteins based on this concept. They constructed a protein variant library and developed a sequence activity model that predicts activity as a function of variant proteins. Hence, the developments of the related technologies make the concept of proteins more popular.

**Table 3. Two patent families citing Smith's SB paper**

| Title | Application Date | Applicants | No of Patent Families | Publication Date | Priority Date | Patent Number | Citations by Patent Families | Citation Context to Smith (1970) |
|---|---|---|---|---|---|---|---|---|
| Method For Producing Novel DNA Sequences With Biological Activity | Sep 30, 1994 | Univ Washington (University) | 2 | Oct 20, 1998 | Jul 17, 1986 | US 5824469 | 64 | No Context. Cited by the examiner rather than the applicant. |
| Methods, Systems, And Software For Identifying Functional Bio-molecules | Mar 3, 2003 | Maxygen Inc, et al. (industry) | 25 | Sep 12, 2003 | Mar 1, 2002 | WO 2003075129 | 50 | Sequence space can be described as a space where all possib1e protein neighbours can be obtained by a series of single point mutations [Smith (1970)]. … Accordingly, new ways to efficiently search sequence space to identify functional proteins would be highly desirable. |

Citations by patent families were accessed via Derwent Innovation Index and counted on May 11, 2018

In addition, the two citing patent families have been cited by many other patent families counted in Derwent Innovation Index by Clarivate Analytics. We identified the fields of technology for each of the 114 citing patent families. Biotechnology is the focus field for both patent US 5824469 and patent WO 2003075129, which yet have been cited by many other technology fields outside of biotechnology. In order to investigate the distribution changes of technology fields over time, we use the year of publication information in the Primary Accession Number, which is a unique identification number assigned by Derwent to each document. This type of "year of publication" is used to analyze the distribution of citing technology fields over time. Figure 7 shows the WIPO technology fields of the citing patent families over time. There are 19 fields of technology for the 50 citing patent families to WO2003075129, which is often cited by biotechnology, organic fine chemistry and pharmaceuticals. And there are 11 fields of technology for the 64 citing patent families to US5824469, which is often cited by biotechnology, organic fine chemistry, analysis of biological materials and pharmaceuticals. So, using the paper-patent citation linkage analysis, we can discover the scholarly article Smith (1970) has influenced what patents and which technology fields. The concept of protein space has influenced number of patented inventions.

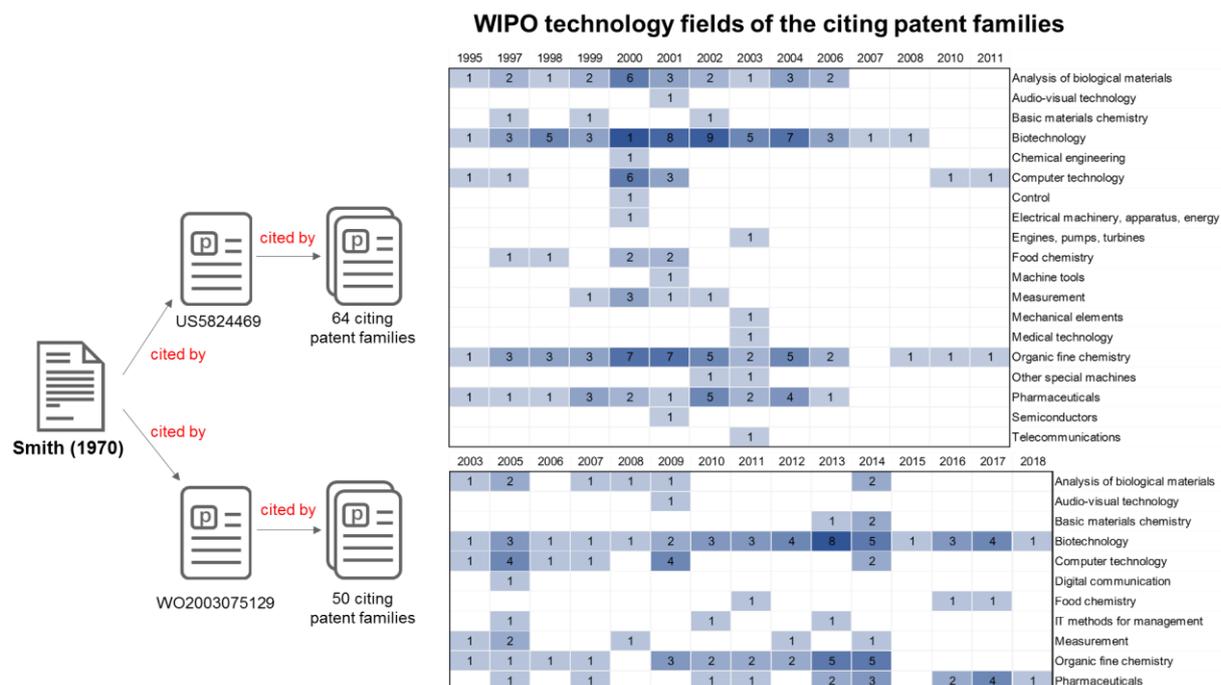

**Figure 7. WIPO technology fields of the citing patent families**

John Maynard Smith, in his 1970 essay, proposed and explicated the concept of a protein space.. As a famous scientist in evolutionary biology, John Maynard Smith's landmark work has laid a very important foundation for follow-up studies in molecular evolution and population genetics. Also, thanks to technological advances, the structural evolution of proteins can be more elaborately studied. This so obviously fundamental discovery has influenced numbers of patented inventions that led to many DNA and protein sequencing products. It appears that the ahead of time discovery firstly encountered negative citations, then patent citations, and finally got widely recognized (Figure 7 and Figure 8).

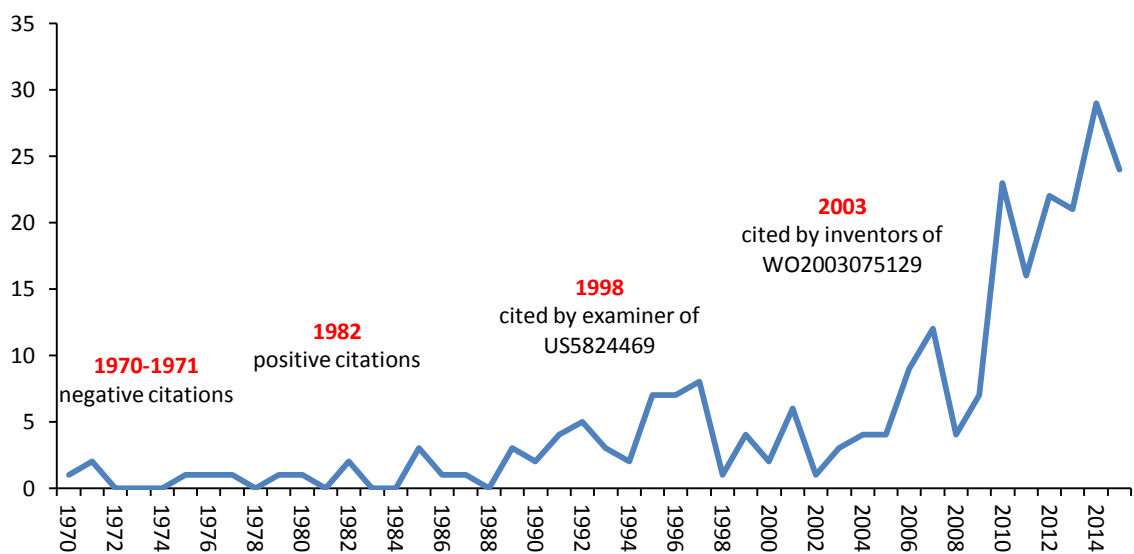

**Figure 8. Citation curve of John Maynard Smith's Nature article**

## 5 Discussion

Rousseau (2018) recently made an important observation that using citations to study delayed recognition is just a—convenient—operationalization of the concept, but that experts may agree on delayed recognition long before this is shown evidently by citations. He discovered a case in which an expert found delayed recognition several years before citation analysis could discover this phenomenon. This leads to the question: How good (or adequate) is citation analysis for detecting premature discoveries? To answer this question, we propose to prioritize the investigation into the lowly-cited papers instead of the most highly cited papers. Combining the frameworks of citation delay and beauty coefficient, we have proposed in our foregoing study a parameter-free index known as *Bcp* index, for identifying under-cited SBs in science, which may indicate possible breakthroughs in an early stage (Du & Wu, 2018). In this study, we further investigate the early signs for predicting delayed recognized knowledge using a comparative study with a control group, i.e., instant recognition papers. It is noted that this paper is a major extension of an ISSI2019 conference paper we published earlier just for a macro level study on Science/Nature articles (Du, Li, Haunschild, Sun, & Tang, 2019). In this paper, in addition to a major improvement of the statistical methods for comparing the patent citing patterns between DR and IR papers, we extended the evidence from a micro level study on a typical case found from the top 10 delayed recognition papers in the above Science/Nature articles. We provide a macro (Science/Nature papers dataset) and micro (a case chosen from the dataset) evidence on the argument that paper-patent citation linkages can serve as early cues for predicting delayed recognized knowledge in science.

The non-patent literature (NPL) cited by patents provide insights into the awakening of delayed recognition knowledge, which may mean that the ahead-of-time discoveries get understood or the transformative potential of research is recognized. In this study, the observed science-technology interactions between fields for Science/Nature papers do not depend on the publishing tendency of two journals. In general, the science-technology interactions for Science/Nature papers are in accordance with those for all Web of Science papers. Here we refer to an annual series technical report and focus on the latest *Japanese Science and Technology Indicators 2019* (National Institute of Science and Technology Policy, 2019). The report analyzed the interactions between technical fields and scientific fields by using the data of Web of Science papers cited in worlds' patents in PATSTAT database produced by European Patent Office (EPO). The science-technology fields were mapped for patent families with the earliest application year between 2007 and 2014, and for papers published during 1981-2014. Such interactions are basically similar fields between science and technology. We can see several major pairs, such as "basic life sciences - biotechnology and pharmaceuticals", "clinical medicine - biotechnology and pharmaceuticals", "chemistry science - chemistry technology", "computer science and mathematics - information and communication technology", and "physics – electronic engineering". For example, Papers in basic life sciences are mainly cited by patents in biotechnology and pharmaceuticals (42.5%), followed by chemistry (25.5%) and biomedical devices (13.1%). Papers in clinical medicine are mainly cited by patents in biotechnology and pharmaceuticals (36.4%), followed by biomedical devices (21.0%) and chemistry (20.3%).

## 6 Conclusion

In this study, we have answered the following two questions , one is the extent of patent citations to papers as early signs of delayed recognition using a comparative study with a control group, i.e., instant recognition papers. The other is the pattern of science-technology interaction between the sleeping science and the technology inducing its recognition.

Using articles published during 1970-2005 in the journals Science/Nature, we conducted a comparative study on delayed recognition with instant recognition papers. Combined with a case study, we found that delayed recognition papers show a stronger and longer technical impact than instant recognition papers. On average, in the more recent years the delayed recognition papers are cited increasingly earlier by a patent. Patent citations seem to play an important role to avoid instant recognition papers to level off or to become "flashes in the pan". We provided further evidence to support the observation made by van Raan and Winnink (2018). This may suggest that early detecting the technological relevance may "prevent" papers from becoming delayed recognition papers. It also appears that the sleeping beauties may firstly encounter negative citations, then patent citations, and finally get widely recognized.

We discovered several pairs of "science sleeping"-"technology inducing", such as "biology-biotechnology/pharmaceuticals", "chemistry-chemical engineering", as well as some trans-fields science-technology interactions for delayed recognition papers, such as "psychology - computer technology/control technology/audio-visual technology", "physics - computer technology", and "mathematics-computer technology". For example, for a given psychological paper which is never cited or poorly cited for several years after publication, once it is cited by a patent outside of its own psychological field, such as computer technology, control technology, or audio-visual technology, it has the potential to become highly cited in the near future. We found that in contrast to the two focused fields (biology and chemistry) for instant recognition papers, delayed recognition papers are rather evenly distributed across biology, chemistry, psychology, geology, materials science, and physics.

## 7 Limitations

It should be noted that the stated conclusions were drawn based on a sample of 400 papers published in two high-profile journals and on case analysis about one paper. We cannot claim that our case study and Nature/Science papers are representative. Our findings are valid within the scope of our study. We have planned in future studies to broaden the focus to other journals.

In a previous study (Du & Wu, 2015), two parameter-free indices, i.e., Citation Speed and Beauty Coefficient *B* were used to identify sleeping beauties (SB) published during 1970-2005 in four high-profile medical journals, i.e., *New England Journal of Medicine*, *Lancet*, *Journal of the American Medical Association*, and *British Medical Journal*. We found that such documents as reviews, clinical guidelines, and books may play a critical role in waking up the delayed recognition in clinical medicine. We may propose a new hypothesis that there are possible different trigger mechanisms behind the abrupt change in the accumulation of citations of delayed recognition papers for different research level. To be more specific, in such cases of basic science discoveries published in Science and Nature for instance, delayed recognition papers with technological importance tend to be 'discovered' and 'awakened' earlier by a patent in an application-oriented context. Whereas for more clinical science discoveries, a long-time never cited or poorly cited paper may suddenly become highly cited once it is cited by a type of consensus-based literature such as review, guideline, and book. It denotes that these cutting-edge concepts, discoveries and/or therapies begin to get understood and accepted by the community. We will validate the possible different pattern in future studies.

We think that neither $Bcp$ nor $B$ is perfect. Imagine the extreme cases where there are three papers, and all their citations are received in the last year of the observation period. But the first one gets 1 citation, the second gets 1000 citations, and the third gets 1000000 citations. Based on the $Bcp$ measure, the three papers have the same "beautiness", but their actual "beautiness" is totally different. Another extreme case is that there are different papers with different total citations but published at the same year, the total number of citations received are all in the last year and the yearly citations after publication till the last year is zero, then they have the same value of $Bcp$ (n-1)/2, n is the age of a given paper. We admit that such cases may exist. But we have checked annual citations for all the 20,000 articles in this paper, and we did not find such cases in our dataset. Nevertheless, the $Bcp$ index can quantify the cumulative citation burst in the entire observing window, while the $B$ coefficient can only measure the annual citation burst before the citation peak. More importantly, $Bcp$ allows comparing the extent of citation delay for any given papers, even if they have different citation patterns.

## 8 Future research

Inspired by our investigations in this study, we propose to combine citation delay analysis with patent & NPL direct citation analysis to identify potential premature and transformative research. The $Bcp$ index proposed by Du and Wu (2018) can be used to identify those under-cited papers that are now happening to be at the sleeping-awakening interface. Afterwards, one could further identify those delayed recognition papers which are also cited by patents. Finally, one could map the structure of the older and long-ignored ideas at both the sleeping-awakening interface and science-technology interface. These ideas and research topics may be the potential origin of transformative research.

Further, inspired by National Institute of Health (NIH)'s Translational Science Search (http://tscience.nlm.nih.gov) and SciTech Strategies Inc.'s approaches for identifying biomedical discoveries (Small, Tseng, & Patek, 2017), we argue that combining text mining based on authors' claims with citation context analysis from citers' comments, one may also discover potential transformative research. It may be possible to use text mining for identifying articles that are regarded by their authors as controversial (they challenge established dogma) or refutation (they disprove previously published data or hypotheses). The author's view can be compared with the citer's view by searching for specific terms (such as "disagree", "contradict", "contrast", "inconsistent", "dispute", ...) in the citation context. It is possible to provide a list of transformative research discoveries from the perspectives of both author's claims and the community's comments.

**Acknowledgments**

The present study is an extended version of an proceeding paper presented at the 17[th] International Conference on Scientometrics and Informetrics, Rome (Italy), 2–5 September 2019. We thank anonymous reviewers for their excellent comments and suggestions, as well as the constructive criticisms from the editor-in-chief. This study was supported by the National Natural Science Foundation of China (Grant No. 71603280) and the National Social Science Fund of China (Grant No. 18BTQ064) and the Young Elite Scientists Sponsorship Program by China Association for Science and Technology (Grant No. 2017QNRC001) and CAMS Initiative for Innovative Medicine (CAMS-I2M-3-018). Data from Microsoft Academic (Sinha et al., 2015) (see also https://aka.ms/msracad) were shared with one of us (RH).

**Appendix Table 1. The earliest 10 citing papers to Smith's paper within 15 years after publication**

| | Authors (Institutions) | Title | Source | Citations | Citation Sentences |
|---|---|---|---|---|---|
| 1 | Spetner, LM (Applied Physics Laboratory, Johns Hopkins University) | Natural Selection versus Gene Uniqueness | Nature, 1970 | 5 | SMITH has compared protein evolution with a popular word game…I would point out, *however*, that it is *not sufficient* to…Smith gives the impression that he is *disagreeing with* Salisbury…In conclusion, after *examining* Smith's argument, Salisbury's contention still seems to stand… |
| 2 | Salisbury, FB (Plant Science Dept., College of Agriculture, Utah State University) | Doubts about modern synthetic theory of evolution | American Biology Teacher, 1971 | 3 | My particular doubt has been published (Salisbury, 1969), scientists have taken their *shots at* it (Smith, 1970), and it has been *defended* (Spetner, 1970). |
| 3 | Butler L (Department of Biochemistry, Purdue University) | Protein structure and properties | Journal of the American Oil Chemists' Society, 1971 | 4 | Obviously *only* an infinitesimal fraction of the latent variability is expressed in actual protein structures. |
| 4 | S.N. Salthe (Department of Biology, Brooklyn College) | Problems of Macroevolution as Seen from a Hierarchical Viewpoint | American Zoologist, 1975 | 15 | Because this model invokes large numbers of selectively equivalent alleles at any locus (Maynard Smith, 1970), it is *only* a contextually richer or more filled-out (but less elegant) version of the neutral allele drift theory. |
| 5 | Olenov, YM (Acad Sci USSR, Cytol Inst) | Evolution of Proteins and Structural Genes | Zhurnal Obshchei Biologii, 1976 | 0 | In Russian, Not accessed |
| 6 | Conrad, M (Department of Computer and Communication Sciences, University of Michigan) | Evolutionary Adaptability of Biological Macromolecules | Journal of Molecular Evolution, 1977 | 16 | The evolutionary **argument** is that evolution proceeds most rapidly when it proceeds through single genetic changes where each genetic type in the sequence has selective value, or is at least viable. |
| 7 | Conrad, M (Department of Computer and Communication Sciences, University Of Michigan) | Mutation-Absorption Model of The Enzyme | Bulletin of Mathematical Biology,1979 | 22 | The argument is that if *this were not the case* each protein would be isolated atop some adaptive peak, with no easily traversable pathways for reaching other adaptive peaks |
| 8 | Mcclendon, JH (School Of Life Sciences, University Of Nebraska Lincoln) | The Evolution of The Chemical Isotopes as an Analog of Biological Evolution. | Journal of Theoretical Biology, 1980 | 6 | Conrad (1978) and Maynard Smith (1970) *have commented on or assumed the same situation* for biological mutations in evolutionary pathways; each member of a pathway is generated by a single mutation and must be adaptively viable enough to permit a second mutation. |
| 9 | Conrad, M (Depts. Of Computer Science And Biological Sciences, Wayne State University) | Natural-Selection and the Evolution of Neutralism | Biosystems, 1982 | 13 | The purpose of this note is to reformulate the rate argument in terms of the protein space *concept first introduced by* Maynard Smith (1970). This is a space of amino acid sequences...For all practical purposes a protein space is a projection of a molecular adaptive surface, *but the slightly different point of view* suggests an especially simple formulation of |

| | | | | | the argument as well as some important features of the bootstrapping process. |
|---|---|---|---|---|---|
| 10 | NODA, H (Univ Tokyo, Fac Sci, Dept Biophys & Biochem) | Probability of Life - Rareness of Realization in Evolution | Journal of Theoretical Biology, 1982 | 5 | Maynard Smith *proposed an interesting idea of a protein space* (Maynard Smith, 1970), within which a track representing the amino acid sequence of a protein crept through over the years of evolution. In his space the amino acid sequence was assumed not to lose function at any point. |

Citations: from publication year till the end of 2018